\documentclass[aps,prl,twocolumn,superscriptaddress,groupedaddress]{revtex4}
\usepackage{subfig}
\usepackage{graphicx}  
\usepackage{dcolumn}   
\usepackage{bm}        
\usepackage{amssymb}   
\usepackage{slashed}
\usepackage{graphicx}				
\usepackage{amsmath}
\usepackage{mathtools}
\usepackage{tikz,pgf}
\usepackage{comment}
\usepackage{asymptote}
\usetikzlibrary{arrows,backgrounds}
\usetikzlibrary{fit,scopes,calc,matrix,positioning,decorations.pathmorphing}
\usepackage[all]{xy}
\usepackage{yfonts}

\newcommand{\bra}[1]{\ensuremath{\left\langle#1\right|}}
\newcommand{\ket}[1]{\ensuremath{\left|#1\right\rangle}}
\newcommand{\Bracket}[1]{\ensuremath{\left\langle#1\right\rangle}}

\begin{document}
\title{Grothendieck's point of view and complexity in the black hole paradox}
\author{Andrei T. Patrascu}
\address{ELI-NP, Horia Hulubei National Institute for R\&D in Physics and Nuclear Engineering, 30 Reactorului St, Bucharest-Magurele, 077125, Romania\\
email: andrei.patrascu.11@alumni.ucl.ac.uk}
\begin{abstract}
These are some speculations on how Grothendieck's point of view and the idea of complexity dynamics  can come together in the problem of explaining the black hole information paradox. They are neither complete, nor final, but can seem like a new direction of research. If read as such they could prove useful to some researchers. The basic idea is that entanglement alone cannot fully account for the information extraction in black hole contexts. Complexity has been proposed as an alternative but remains a vague concept. I employ Grothendieck's point of view to expand the idea of entanglement entropy to a categorical context in which the objects (states) and their maps are considered together and the map space has additional topological and geometric structure that intermingles with the object set of the category via Sieves, Sheafs, and Toposes. 
\end{abstract}
\maketitle
\section{Introduction}

\par The black hole information problem is finally a communication problem [1]. We have to understand a set of rules of communication, given at least general relativity and quantum computing. These two may or may not be the sole ingredients. They encode basically two concepts: the locality and causality structure for information transmission, as expressed in special and general relativity, and the rules of information encoding, sometimes in a non-cartesian (i.e. non-separable) manner, and most of the times in a global manner (superposition of probability amplitudes), in quantum information. The key word here is "transmission". We are used with the idea that information is a set of signals that get transmitted from point $A$ to point $B$, in the best possible case, by massless gauge fields, aka on a gauge connection with curvature. We can think also in terms of flat connections with topological structures, and we get to Chern-Simons terms. Those are interesting in the context of topological invariants etc. but are not directly solving the information problem. 
At least most people would agree that the information problem must find its solution in some form of quantum gravity, where quantum information and particularly quantum entanglement play a significant role [2], [3], [4], [5]. 

\par In mathematics the idea of information extraction is seen in a rather different way. While in physics and engineering we consider information transmission as in, following a connection between fibres, or, following a rule in which the information about an event in one region will traverse a region of space and will be consistently moved to another region of space, in mathematics we have most of the time some form of inference of results from a certain underlying structure. 
\par Among the important results of the 20th century are the theorems by Grothendieck, and among them, many are based on a principle called "the relative point of view" [6], [7], [8]. This idea means, roughly said, that instead of looking at an object in a category or at one single category to derive its properties, we can do the same by looking at all possible maps between that category and another category or object considered in some sense as a terminal structure in a chain or as having some universal properties [9]. The ideas of the classifying spaces are among the first that would materialise this concept [10], but a truly powerful idea would be to apply this to fibre bundles over spaces. Grothendieck's construction in its most basic form does precisely this: if $X$ and $Y$ are spaces and $A$ and $B$ are vector bundles on top of them, then all maps from $X$ to $Y$ determine an inverse set of maps from $B$ to $A$, in the sense that they constrain and define the connections on the former by all the maps defined between the spaces themselves. 
\par 
This may sound abstract at first. The whole idea is to determine the vector bundle structures by means of pullbacks on the spaces, and that would include the connections. What we need is some prior knowledge about the connections and it seems that is precisely what we obtain from the maps between the spaces and the inverse maps on the bundles. With such prior knowledge, the actual transmission of information becomes secondary, as information regarding the final point, after crossing the fibres via a connection, is already encoded in the geometry and topology of the maps between the space we work in and another benchmark space. 
\par Another way of looking at it is considering the idea of an information channel required to transmit bits of information from one side to the other. The amount of information that needs to be transmitted is encoded in the informational entropy, which has an identical formal expression with the thermodynamical entropy. We can interpret this in the following way: If an information channel has two ends, $A$ and $B$, then for $B$ to know about an event in $A$ one needs to transmit the information describing the event in $A$ to $B$. However, as is the case in quantum information, when shared entanglement generates a global state that encodes the information about the system, say $\rho_{AB}$, not all information about $A$ needs to be transmitted to $B$. Some of the information is already accessible in $B$ due to the global nature of $\rho_{AB}$. The entropy measures a degree of randomness, namely exactly what cannot be inferred by $B$ without information being transmitted to $B$. Now, this way of thinking is based on the idea that the underlying spacetime (or the fibre bundle used to define the connection between the two points) is uniquely defined. If however we think in terms of a categorical approach, we may study the bundle with the connection linking the two ends by using all possible maps that would link the current bundle with a terminal bundle. Additional information about the events in $A$ and $B$ will provide non-trivial structure to those maps, maybe some form of obstructions or non-separabilities of maps, etc. Therefore the map space will contain the same information that is available in the space where $A$ and $B$ reside, but that information will be retrievable without following a specific connection between $A$ and $B$.
\par The idea behind ER=EPR is that an entangled state between two regions on the boundary is dual to a wormhole geometry connecting the two boundary regions in the bulk. Geometrically we think of the bulk space as a region with one additional dimension, through which we can define wormhole geometries that would not be visible on the boundary. This represents a geometric extension of the system to be studied. A categorical extension a la Grothendieck, which is what I am proposing in this article, is one in which we do not extend only towards an additional dimension, but one in which we extend our space structure (or fibre structure) by considering also all possible maps to some pre-established and well known benchmark structure. Instead of saying entanglement amounts to "travelling through a bulk wormhole", we also add that information about the two separated regions can be extracted from all the ways our spacetime (or bundle) can be mapped into some benchmark structure. Grothendieck showed that the map structure can have all the properties of a space, with geometry, topology, etc. and can even be endowed with a Hilbert structure if need arises. 
\par
Thinking at how and where this could be useful in the black hole paradox I came to an interesting realisation: when a quantum entangled or even a coherent state, starts interacting with the environment, its entanglement is spread thin due to its spreading in the environment. This is part of the process we call decoherence, and it is usually regarded as a problem in quantum computing. After all, what we want to make use of, namely quantum entanglement for a controlled system, is hard to maintain given any interactions with an environment. This is sometimes considered in the problem of growing complexity, or in the process of the so called black hole scrambling. It appears to be precisely this type of complexity that is given by the set of maps linking the system with the "terminal object" or the "universal object" in the classifying space description. That mapping structure, produced by what we call "decoherence", is precisely what could offer us an inverse map (a pullback) on the fibre bundles defined on our initial space and on the space where these maps lead to. The result would be inverse maps that would refine our concept of a connection, given the information that is spreading out in the environment. We would get a "predictive model" for our connection, which is exactly what we want, namely, to ensure either enough entanglement or enough informational contact between objects that would appear otherwise causally disconnected. 
In a sense, it is not the usual causal communication that establishes entanglement between the in-falling object in the black hole and the emitted radiation. However, there would be a way in which information about the object can be retrieved from the possible maps between the in-falling object and the black hole's horizon region where particles are being created. 
I wonder if this mechanism could play the role of a holographic map and a potential categorical interpretation thereof? 
This approach, due to Grothendieck, is in many ways similar to the fact that entanglement is spreading in the environment when interactions with the environment become possible. 
The idea leading to this conclusion has had several incarnations up to now, ER=EPR (the duality linking entanglement or the Einstein Podolski Rosen paradox to the spacetime geometry of wormholes, the Einstein-Rosen bridges) being one of them. ER=EPR is more of a descriptive duality. Thermofield double states are formed by entangling two copies of a conformal field theory such that when we trace out one of the copies we obtain a thermal density matrix for the other. This is equivalent to an eternal black hole in AdS. In particular the two sides of the eternal black hole dual to the thermofield double state are geometrically connected by wormholes and the volume of such a wormhole has a dynamics governed by a growth time proportional to the exponential of the number of degrees of freedom of the boundary theory. 
As we know that simple entanglement is not sufficient to describe the quantum information paradox, this value for time, which defines a scale much larger than the usual scales involving entanglement entropy, is quite interesting. The mutual information saturation time or the scrambling time provide us with a scale of the type
\begin{equation}
t_{*}\sim \frac{\beta}{2\pi} log (S)
\end{equation}
where $\beta$ is the inverse temperature and $S$ is the entropy of the black hole.
This scale is dwarfed by the wormhole volume growth time scale. This time scale has been connected to a loosely defined "quantum computational complexity" of the boundary state. A lot of work on quantum circuitry has been performed in order to understand the details of such a connection [23], [24], [25]. There have been two principal measures for this complexity that have been postulated:
On one side we had the "complexity=volume" postulate where the complexity is defined as a co-dimension-$1$ volume of the maximal spacelike slice that connects the two CFTs involved in the TFD state. On the other side we had the "complexity=action" conjecture postulating that the dual is the gravitational action of the bulk region bounded by light sheets. 
However this notion of complexity is incomplete in various ways. Mostly the problem appears because one tries to describe complexity by means of state-bound information. I think here is where Grothendieck's point of view can play an important role. 
It is not only the information related to one reference state and all the states it will be mapped into, by means of the  various possible circuits, that encode the "complexity". The geometrical and topological structure of the maps themselves can reveal information about the system. From a categorical point of view it is never only the object that plays a role, but also the mathematical properties of the maps between the objects inside a category and even the ways the category can be mapped by various functors into other categories. Basically the searches are for an "optimal circuit". I think this approach is insufficient. Yes, there are many (possibly infinite) possible circuits and maps, but their properties can be described quite generally and out of those properties we can infer more relevant details about what complexity is meant to be. 
Moreover, the maps between those objects have a geometry, and even a topology that may encode separable and non-separable information, therefore there must be a concept of entanglement entropy describing the information that can be globally encoded in the mapping structure. This also means that there can be information encoded both locally and globally in the geometry of the maps.
\par To be more formal on this aspect, let there be a map that connects an initial and a reference state
\begin{equation}
M:\ket{x}\rightarrow \ket{y}
\end{equation}
then the map $M$ may be indexed by some parameter $i$ resulting in the notation $M_{i}$ which encodes all possible maps. This set may be infinite, but even so, it has a topology and a geometry, given the algebraic structure obeyed by $i$. Therefore there will be a map $\phi$ connecting those maps, in the form of 
\begin{equation}
M_{i}\xrightarrow{\phi}M_{j}
\end{equation}
Now, thanks to Grothendieck, and previously to the Gauss-Mannin equation for a more simplified case, we can associate a geometry for these maps between states $M_{k}$ depending on the maps $\phi$ between them. In particular we could define a connection on them, this connection allowing us to move between the various geometries of the maps in a consistent manner, in the same way we move between fibres in a bundle, and then we could define a topology of those maps, and hence some form of global data between them and move between maps of different topological classes in the same way. We can see that the notion of a connection admits quite an amazing level of generalisation, mostly thanks to Grothendieck's broad generalisations.  
Both the maps and the global data on them contain information about the initial and final state as well as about their informational characteristics. The total complexity associated to it will be far larger by doing this, but it will be more complete, and not necessarily harder to deal with. Not always more complexity is worse, sometimes it can even prove to be better. 
This global data then can be related by means of a Grothendieck topology to other data, in the same way in which we can define pullback maps between fibre bundles over spaces connected by push-forward maps. 
Then comes information and its separability. If information about mapping is to be non-separable as defined on the mapping spaces, this amounts to an obstruction to cartesianity, but as opposed to the previous essays on this subject, not in the space of states, say, as in the space of wavefunctions, gauge fields, etc. but instead in the space of maps between states. This is obviously a categorical approach, just not the usual one, because the maps become maps between quantum states, and they have a peculiar fibre bundle interpretation themselves. This makes us come again back to Grothendieck's construction. 

\section{higher quantum mechanics}
The rules of quantum mechanics have, by themselves, universal applicability. We start with the observation that in quantum mechanics, non-realised but potentially allowed states (allowed by various symmetry constraints, etc.) do have the possibility to contribute to the statistics of measurements we perform. In that sense, particles that would only exist (or be realised) at very high energy, do in fact contribute to the lower expansion orders through the inner loops, over which we need to integrate while keeping the particles entering those loops "off shell". The "off shell" nature of those inner loops encodes the fact that their contribution to the amplitudes doesn't occur through a physical realisation, but only through their potential existence at higher energies. In an equivalent manner of speaking, we have to sum over all possible paths in order to obtain Feynman's path integrals. Due to this fact, quantum mechanics has the possibility, to a certain extent, to give us insights into effects that are not usually perturbatively accessible. In the Feynman diagrammatic expansion, the loop diagrams do contain information that can be extracted about particles that would only be real and on-shell at far higher energies. The path integral approach gives us information about the global structure of the manifold over which the photons or electrons move, encoded in the amplitudes of probabilities used to calculate the probability of particles reaching the measuring screen in different positions. Therefore quantum mechanics clearly has the potential of accessing information about the manifold, that is globally encoded, even though it is not accessible to a perturbative observer. This is extremely visible in the context of the SU(2) gauge anomaly, where an otherwise apparently well defined theory turns ambiguous due to the fact that the higher homotopy of the SU(2) group can in fact be probed by quantum mechanics, turning the quantum partition function ill defined.
\par There are at least two ways in which we can make better use of this fact. To some extent I suspect those two approaches are equivalent. 
First, it is probably well known (as the hierarchy problem) that we cannot fully eliminate all effects of the UV couplings and degrees of freedom. In fact, I showed (ref. [17]) that a low energy effect of the string theoretical T-duality could have a role in explaining the hierarchy problem. Different approaches by different authors [19, 20, 21] support the view that T-duality may have a role in understanding the hierarchy problem. The scale decoupling we are supposed to achieve by means of renormalisation and renormalisation group calculations is probably only partial, and there seems to exist a lot of mathematical and physical structure that can only be obtained by means of probes that are not elementary particles. Indeed, a string can "detect" properties of its surroundings that cannot be detected by a particle. However, strings have been obtained by postulating that objects at high energy gain (or fundamentally have) an additional dimension, in particular length, as opposed to the over-simplified points. The procedure of quantising strings is decently well understood, but as a prescription, it still first assumes classical strings as a point of start, and then applies a series of procedures that amount to the construction of a consistent quantum theory for such objects. 
This approach limits the scope of our thinking to a realm where strings are indeed having an additional dimension (therefore a realm of extremely high energies) and doesn't allow us to look at lower energies, at least not easily, if we start with a string theoretical point of origin. While we can imagine a series of high energy - low energy dualities, they are simply point-wise probes of probably a vastly more complex physical and mathematical structure. 
\par I believe that the assumption that low energy effects do not present structures that could be associated to something that only extended objects could detect is an oversimplified approach. However, at low energies, the rules of quantum mechanics are forced to rely on either point-like particles (as is the case in the first quantisation), or on the relativisation of the idea of particles (The number of particles operator is not truly a diffeo- (or gauge-) invariant observable, and even in special relativistic quantum field theory it is not a conserved quantity, therefore allowing for creation and annihilation of particles). In both those cases we try hard to force our quantum theory to accommodate localised particles and localised effects. Problems are bound to emerge, and in fact we have the Reeh Schlieder theorem revealing some of those "problems". 
Now, I propose here a method by which we can expand the usual approach to quantum mechanics, to a construction that is categorical in nature, and hence takes into account higher form effects as well as their effects on lower energy physics. However, this expansion does not rely on the quantisation of some intrinsically 1-dimensional (or higher dimensional) object. Let us not forget that the quantisation of D-branes is not truly fully understood. Instead, I assume I do not know from the beginning what the effects of some non-point-like high energy structure would be at low energies, or even if such a structure exists at all, but I will focus on the ability of quantum mechanics to probe the complicated structure of the manifold on which it operates. 
\par In a sense I consider this property a foundational property of quantum mechanics, or maybe, better expressed, an axiom: a quantum theory is a theory that can, by means of alterations of probabilistic outcomes (via probability amplitudes for example), give access to the full topological and geometrical structure of the manifold or fibre bundle it operates upon, access that would otherwise not be accessible due to energy, coupling, or distance limitations. Other theories can reveal such extended global information by other means. For example general relativity by means of spacetime curvature and (co)bordisms. I speculated in ref. [22] that theories that share this ability of probing beyond the perturbative regime are linked by some more or less manifest dualities. I know of three such instances: quantum mechanics, general relativity, and neural networks. Understanding the connection between them is quite speculative. However it is worth mentioning that the fact that any tools used to describe the same global structure should agree on the properties of the global structure and its local effects, amounts to relations between various aspects of those tools, which ultimately share a common goal. It is therefore interesting to remind the reader that the idea of probing topological structures by means of Morse theory connected the field of topology to that of differentiable functions on a manifold. 
\par In physics, it would be interesting to understand the connection of, say, quantum mechanics and neural networks, as they both have the ability of probing global effects that are not detectable strictly locally / perturbatively. Some of the properties of such structures are not encoded directly in the structure itself, but instead, in the properties expressed by the maps that connect that structure with another structure, either chosen as a benchmark, or encountered "in the wild". A simple example would be a torus. We can realise that a torus is different from a disk by trying to transform one into the other and encountering a series of divergent structures that become "obstructions" to such a transformation. However, it is indeed possible to transform a torus in a disk if we change the inner structure of the point by changing the coefficients of the cohomology we use to detect such a topological feature. Some cohomology theories may simply be blind to the particular distinction between a disk and a torus, but could, at the same time, potentially be sensible to twists or other features [18]. 
In any case, to be able to discuss quantum effects taking truly into account all available information, we need to take into account the structure of the maps that transform our underlying manifold (or even bundle) into another manifold or bundle. It is always fascinating to discuss about theories that have to take into account "everything possible", because more often than not, "everything possible" may just as well be more than what we imagine at a given time. Quantum mechanics implies the inclusion of "everything possible" in the amplitude description. It so happens that the maps between our underlying structures amount to maps between quantum states defined over such structures. Such maps will have properties similar to standard quantum mechanics, for example, could be entangled, hence inseparable, could generate superpositions, etc. 
\par In a sense, I am trying to explore the following analogy. Consider a quantum gauge field theory. Expanding the gauge theory from the simple 1-form field, 2-form field strength representation to a higher gauge form structure we obtain string theory, the associated string and D-brane gauge fields (Kalb-Ramond fields, etc.).
The quantisation of the string side of the story is somehow assumed, although not at all certain due to problems in quantising higher form objects, D-branes, etc.
What if instead of thinking in these terms, we would be thinking : what kind of quantum mechanics do we need in order to take into account or probe structures that are not detectable only by point-like probes? Instead of expanding the objects and their gauge interactions to higher forms, what if we expanded quantum mechanics in order to automatically take into account higher form and categorical effects?
Indeed this is possible, and is what I am doing in a simple example in the following chapters of this article. 
Instead of postulating the existence of higher energy (hence lower length scale) extended objects, what if we re-designed our low energy quantum mechanics to be able to probe even low energy but higher form / higher categorical structures? 
\par Quantum mechanics seems to be the best tool that would be capable to do that. As said before, quantum mechanics has the ability to statistically probe the far regions of the permissible manifold (or of the setup of our experiments) that are not accessible to a standard perturbative approach. 
\par As a toy experiment, it would be interesting to present the following situation. We all know of the Bohm Aharonov experiment. Basically a thin coil with an induction field inside it, $B$, is placed in the middle of a platform, with electrons being transmitted around it. The induction doesn't reach the exterior of the coil, being perfectly well confined in the inside. However, the field $A$ does exist also outside, and due to the fact that the coil induces a change in the topology of the region accessible to the electrons, this will appear as an compensatory shift effect on the position of the interference pattern of the electrons on the screen placed at the end of the platform. This is of course well known. However, let us extend this experiment slightly. Instead of having a simple two dimensional platform and a single topological hole in the middle, let us assume we can send electrons all around in the volume of space between the detector and the electron gun. Instead of a simple coil, let us have a toroidal coil. The toroidal coil will represent a more complex topological feature now, but it will still not be sufficient. Let us consider now the toroidal coil with a knot on one side. Compare this experiment with the situation in which the knot is replaced with a tangle. A tangle is a situation in which the knot doesn't quite close, it leaves the two sides open, yet, in a good approximation, the torus will be almost closed (see figure). In mathematics we have an instrument called Khovanov (co)homology which is expected to emerge as some form of observable of a 4-dimensional topological quantum field theory in the same way in which the Jones polynomials emerged as observables of the Chern-Simons topological theories. Jones polynomials do indeed appear in the decategorification of Khovanov's cohomology. Jones polynomials classify knots, but Khovanov cohomology is a far more powerful, albeit not fully understood, invariant. In fact, the graded Euler characteristic of the Khovanov homology is the unnormalised Jones polynomial. The ability of deforming knots into equivalent knots leads to an additional structure added to our experiment. We get equivalent structures and hence some form of interference pattern of sorts. We can even project our knot into a toroidal coil, however, there is no simple deformation that can link a knot to a tangle, as the tangle is a section of a knot, and it is fundamentally not closed. If however the tangle does basically link the two sides of our torus, it will be physically almost similar to a situation in which we simply have a torus. Unless, of course, we can also probe the space of possible maps between the various structures. If our quantum mechanics becomes sensitive to such maps, we can easily determine that the maps will have obstructions into allowing themselves to be deformed continuously between mapping two knots and mapping a knot and a tangle. A quantum mechanical construction sensitive to maps between manifolds or bundles would therefore detect such obstructions and express them in the form of some detectable discrepancy from the case in which we deal with a closed knot. 
What the (topological) quantum field theory associated to Khovanov cohomology could be remains a mystery. But as a technical tool that can be important to physicists, it would be interesting to design a form of quantum mechanics that would, without assuming that high energy structures contain higher form fields and higher gauge structures, be able to detect the distinction between a knot and a tangle by being sensitive to the separability of the algebraic structure of the maps between states. 
\par This rather different approach has an impact on understanding various aspects of quantum gravity and black hole physics. As is the case close to the horizons of black holes, but also in various condensed matter systems, we do not directly encounter strings or other extended objects. However, various obstructions occurring in higher form physics do have an impact on various measurable quantities. In particular, the physics around the black holes is far from being stringy by itself, but is nevertheless expected to present effects that cannot be explained without access to some higher form quantum mechanics. This situation seems somehow unpleasant, as we should not yet require string theory to solve the black hole information problem, simply because the involved energies are far from what is expected string theory to be relevant for, but we do expect various non-local effects or at least some form of quantum correlations that are beyond what the usual understanding of entanglement would imply. 
\begin{figure}
  \includegraphics[width=\linewidth]{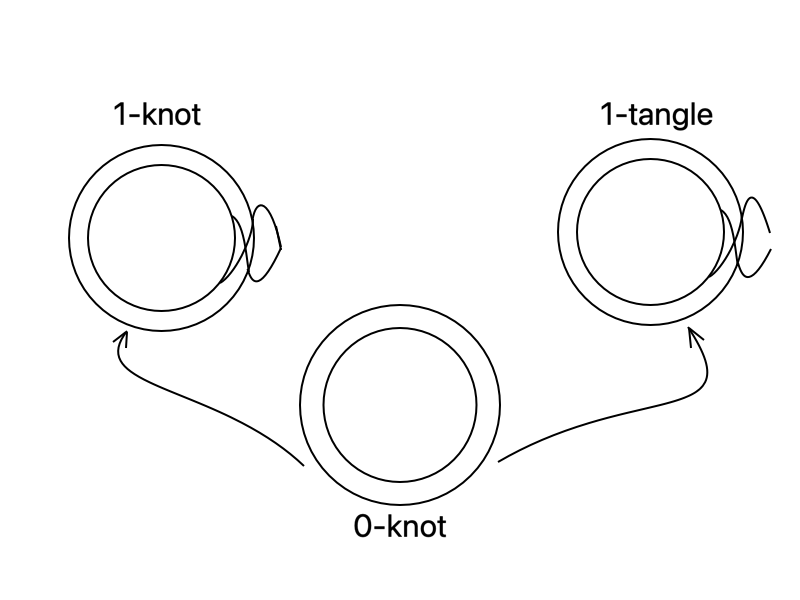}
  \caption{at the bottom of the figure: a zero knot, or a torus coil. The one-knot is one of the 3-dimensional embeddings of the zero-knot. On the right side, a one-tangle, which cannot be obtained directly from the knots but is expected to be described by means of categorifications involving Khovanov cohomology. A quantum theory as we understand it now will have a hard time detecting such a feature. However, a quantum theory that includes additional algebraic structure for the maps between states will detect such a distinction by means of a transition between separability and non-separability within the space of maps between states. In order to construct a full quantum theory that can indeed detect global aspects of its underlying manifold, such a quantum theory should be extended in a categorical way.}
  \label{fig:gauge1}
\end{figure}
\par As opposed to string theory, which I call a "postulative" theory, in the sense that it postulates a certain type of extension into the UV domain, namely, it postulates that higher energy objects will be strings having a fundamental new dimensions, a quantum theory at low energy scales that would be able to probe into the geometrical and topological properties of the maps between states, hence would imply a certain categorification, would be, what I would call, an "explorative" theory, in which we do not know the effects of extended objects in our energy domains, but if existent, we would be able to probe them and detect them as such. Instead of creating the extension at the level of the microscopic structure of the objects, we can create it directly in quantum mechanics, encoding the possibility to determine more structure by extending quantum mechanics to a theory capable of probing not only in the non-perturbative structure of the underlying manifold but also into the (potentially non-perturbative) non-separable structure of the map-spaces. 
\par As a historical note, it is worth mentioning that attempts to carry the stringy effects down to the low energy effective theories have been seen before. One of those attempts is the so called double-field theory[29], [30], [31], a theory that includes additional degrees of freedom (namely a doubling of the degrees of freedom) of the low energy theories in order to include winding variables. As strings can in principle wind around cycles in higher dimensional space compactifications, it was considered that adding those additional degrees of freedom and treating them in a field theoretical way would amount to obtaining stringy effects in effective low energy field theories. In particular, it was attempted to encode T-duality as a symmetry of the lower energy effective theories. The approach encountered however some major difficulties and never became widely used. My approach here is fundamentally different. Even with basic quantum mechanics, I can consider geometrical and topological structure associated to the spaces of maps, while no additional geometrical structure is needed for the base spacetime or the bundle used to describe quantum mechanics. As stated previously, this expansion does not go into a higher geometrical space, but instead in the geometrical and topological properties of the space of maps. Also, I never postulate additional properties at the high energy end of the theory, in the form of a string theoretical continuation. I remain agnostic towards the higher energy continuation and consider only how higher form effects could be detectable by quantum mechanics if they appeared to have an impact on low energy physics. While the depths of the black hole may indeed be described by string theory, dealing with informational problems at the horizon of black holes doesn't immediately require such string theoretical techniques, but instead a different approach to more conventional physical structures. 
\section{Thought experiments}
As mentioned in the previous section, if we admit as an axiom of quantum mechanics that it can provide us with non-perturbative or non-local information via a (pre-)probabilistic approach, we have to see where such an approach can lead us. What other theories do depend strongly on non-local features? As said before, I know of general relativity, and as an extension, quantum gravity, as being one such theory, and then I know of the emerging field of neural networks that has this property. In general relativity and quantum gravity such non-local properties appear due to the fact that gauge invariant observables cannot be local, hence even in the cases in which we have matter fields coupled to gravity, we need to append dressing operators to the usual matter fields to make them properly diffeo-invariant. This suggests that general relativity is itself also a theory that allows us to probe non-local features or, better said, to probe into the global structure of the manifold where we placed our theory. It is important to make a distinction between a theory and the tools we use to extract information from it. Indeed, while both quantum mechanics and general relativity are essentially capable of taking into account information about the global structure of our manifold, the methods we employed until now are not so sensitive. Perturbative expansions are by definition bound to regions around a bottom of some potential, lattice methods are bound by shaky approximations to the continuum limit, etc. It is interesting to look for a bit at the connection to neural networks and how they can take into account global features, as indeed they do. We can indeed imagine a neural network simulating a set of agents that act in various ways, their policies being optimised such that some global feature is probed or obtained. With this, it is easy to train systems to classify various objects, taking into account their global features, a capability that couldn't emerge from a strictly local approach. As we know (or can read about) the inner workings of the neural networks and of establishing agent policies by means of neural networks, it is probably more interesting to present the thought experiment that made me consider the solution presented in this article. Let us assume that we are probing global features of an underlying manifold with two different approaches that are based on different theories both capable of obtaining such global information. Say, one of them is quantum mechanics which enables us to determine global properties by means of quantum phases and the potential non-classical correlations of the measurement outcomes. We become sensitive to non-perturbative or non-local effects by means of the effects that such remote objects have on the sum over amplitudes we obtain at our energy. For example in the process of Feynman diagrams calculations, often, new physics appearing somewhere at higher energies is inferred by the modifications such new physics particles induce on the amplitudes due to the integration over inner loops that must take such high energy (certainly off-shell) contributions inside the loops into account. 
On the other side, let us try to describe the same underlying manifold, with the same global features, by means of another theory that can reveal such global structure. Say, neural networks. Knowing that the underlying manifold is the same, the two theories must be bound by correspondences between various of their features in the same way in which properties of the Morse functions are bound to topological features determined about the space in which they reside. To make this possible however, we cannot rely on the theory defined only in the way it is, in a decategorified way, but we also must include all possible maps the respective theory involves in making non-local predictions. In this article I simply show one particular case in which the geometric and topological properties of such maps can be of importance. I also study the non-separability of the spaces encoding such maps, leading to what I call a "higher entanglement". 
\section{Thermofield double states, entangled conformal field theories and the problem of black holes}
This article is mainly about a new understanding of non-perturbative effects in general, by looking at how quantum mechanics can give us access to the non-perturbative realm. As stated previously, if one considers the ability of quantum mechanics to give us access to non-perturbative information in a probabilistic manner as a foundational axiom of what quantum mechanics means, we can expand the applications of quantum mechanics in various directions. As stated, quantum mechanics implies taking into account all possible intermediate states. This is either done in a Feynman perturbative expansion by integration over inner loops momenta in Feynman diagrams, or by means of the Feynman path integral, taking into account all possible paths connecting the initial and final state in an experimental setup. It is probably important to understand why a traditional probabilistic approach would not suffice. Indeed, if we just calculate probabilities of events in a classical context, by means of the usual Bayesian approach, we obtain results based on effects that could possibly happen and do in fact happen, giving single outcomes. In quantum mechanics probabilities are obtained from a pre-probabilistic amplitude that is used, allowing for outcomes that could possibly happen to interfere physically, before they are being realised. Moreover, all such possible outcomes must be taken into account to produce the proper amplitude from which the probability is calculated. Therefore, as opposed to a classical Bayesian approach, the probability amplitudes do actually probe the non-local and non-perturbative regions of our manifold, because such "remote" states or configurations do actually play a physical role in calculating the amplitudes of quantum mechanics. 
\par Another quantum property is that indeed, all possible intermediate states need to be taken into account, however, if that is the case, the categorical expansions must also be taken into account, hence all possible maps between configurations must be included, or the quantum interpretation will not be complete. 
The application of this insight is not exclusive to quantum gravity or the black hole paradox. In fact I am in the process of preparing a paper where this view is analysed from a renormalisation group point of view. 
\par However, because I mentioned some more obvious applications to quantum gravity, I will explain here some aspects of our understanding of the most popular dualities of quantum gravity (AdS/CFT and ER-EPR) that are more relevant to this article. 
First, let us start with a thermofield double state. If we are given two copies of any quantum mechanical system, we will call a thermofield double state (or $\ket{TFD}$) the following unique pure state
\begin{equation}
\ket{TFD}=\frac{1}{\sqrt{Z}}\sum_{n}e^{-\beta E_{n}/2}\ket{n}_{L}\otimes \ket{n}_{R}
\end{equation}
where we have the system consisting of subsystems with $\ket{n}_{L,R}$ the energy eigenstates of the individual subsystems. This is an entangled pure state of the full system, given that each of the two copies is in the thermal state of a density matrix with temperature $\beta^{-1}$. 
\par This state is important in quantum gravity as it is usually understood as a dual state to an eternal black hole [32]. Now, if we are dealing with AdS black holes, then, if we apply some simple perturbative couplings on the two sides of such an eternal black hole, we may obtain a traversable wormhole. This is important because it allows us to send probe particles through this structure and to determine the response of the structure to the probe particle by analysing the probe particle when it emerges. Such an eternal AdS black hole is dual to the thermofield double state that encodes the system formed by the two boundary CFTs of the AdS black hole. Let us see how such a state can be constructed, and while doing that, further look into its properties. For that I will follow ref. [26]. 
We may start with two identical quantum systems and start some interaction between them such that the ground state of the overall system is precisely a thermofield double state. 
We define the thermofield double state as
\begin{equation}
\ket{TFD}=\frac{1}{\sqrt{Z}}\sum_{n}e^{-\beta E_{n}/2}\ket{n}_{L}\otimes \ket{n^{*}}_{R}
\end{equation}
where the second state $\ket{n^{*}}$ is related to $\ket{n}$ by means of an anti-unitary operator $\Theta$. Consider an operator acting on the left theory, $\mathcal{O}_{L}$ and $\mathcal{O}_{R}$ the corresponding operator on the right theory. We can employ these operators to annihilate the thermofield double state 
\begin{equation}
d\ket{TFD}=0
\end{equation}
An operator that would do this would be
\begin{equation}
d=e^{-\beta(H_{L}^{0}+H_{R}^{0})/4}(\mathcal{O}_{L}-\Theta \mathcal{O}_{R}^{\dagger}\mathcal{O}^{-1})e^{\beta(H_{L}^{0}+H_{R}^{0})/4}
\end{equation}
where $H_{L}^{0}$ is the original Hamiltonian of the left theory, $H_{R}^{0}$ is the original Hamiltonian of the right theory, $\mathcal{O}_{L}^{k}$ is any operator in the left system, and $\mathcal{O}_{R}^{k}$ is the same operator in the right system. Such a Hamiltonian has the exact thermofield double state as its ground state, and hence a variational approach on it will provide us with the thermofield double state. 
This can be seen if we look at the matrix elements of the two terms in $d$ in the energy eigenbasis. We can clearly see that 
\begin{widetext}
\begin{equation}
d\ket{TFD}=\frac{1}{\sqrt{Z}}\sum_{ij}e^{-\beta(E_{i}+E_{j})/4}((\mathcal{O})_{ij}\ket{i}\ket{j^{*}}-(\mathcal{O}^{\dagger})_{ji}^{*}\ket{i}\ket{j^{*}})=0
\end{equation}
\end{widetext}
where $(\mathcal{O})_{ij}=\Bracket{i|\mathcal{O}|j}$. 
We can now define a hamiltonian associated to thermofield double states, that will be formed out of such operators
\begin{equation}
H_{TFD}=\sum_{i}c_{i}d^{\dagger}_{i}d_{i}
\end{equation}
with $c_{i}$ positive numbers. A state that is annihilated by the $d$ operator built from $\mathcal{O}_{1}$ and by the $d$ operator built from $\mathcal{O}_{2}$ will also be annihilated by the $d$ operator built from their commutators. Therefore we have a set $\mathcal{A}$ of elements that generate all the operators in the quantum field theory by commutation algebra, and hence we can define our hamiltonian as
\begin{equation}
H_{TFD}=\sum_{i\in \mathcal{A}}c_{i}d_{i}^{\dagger}d_{i}
\end{equation}
The ground state of such a Hamiltonian is unique. The proof of ref. [26] is useful. We can linearly map the Hilbert space of the double theory to the space of operators of the single sided left theory, provided a certain choice of the anti-unitary operator $\Theta$ is made, by defining a map 
\begin{equation}
\mathcal{M}:\ket{n}_{L}\otimes \ket{m^{*}}_{R}\rightarrow \ket{n}\otimes \bra{m}
\end{equation}
and we can find the set of operators that lie in the kernel of the super-operators related through this linear map to $d$
\begin{equation}
\mathcal{D}_{i}=[\mathcal{O}_{i},\cdot]_{\beta}
\end{equation}
for all $i\in\mathcal{A}$ where the commutator above is defined by 
\begin{equation}
[\mathcal{O},\mathcal{Q}]_{\beta}e^{-\beta H^{0}/4}\mathcal{O}e^{\beta H^{0}/4}\mathcal{Q}-\mathcal{Q}e^{\beta H^{0}/4}\mathcal{O}e^{-\beta H^{0}/4}
\end{equation}
The sole operator that commutes by means of this commutator with all operators in $\mathcal{A}$ is the identity 
\begin{equation}
\mathcal{I}_{\beta}=e^{-\beta H^{0}/2}
\end{equation}
If we transform back this operator by means of the inverse map $\mathcal{M}^{-1}$, this operator corresponds to the thermofield double state. This means that $\mathcal{I}_{\beta}$ is the unique ground state of the super-hamiltonian
\begin{equation}
\mathcal{H}_{TFD}=\sum_{i\in \mathcal{A}}c_{i}[\mathcal{O}_{i}^{\dagger},[\mathcal{O}_{i},\cdot]_{\beta}]_{-\beta}
\end{equation}
This Hamiltonian also has a positive semi-definite spectrum and a unique ground state with $\mathcal{H}_{TFD}=0$ for the state $\mathcal{I}_{\beta}$. 
Of course there can be various operators that can annihilate a thermofield double state, so the Hamiltonian can be constructed with any of those. As long as the properties mentioned above are satisfied this is not of fundamental importance. 
Therefore following reference [26] we derived a Hamiltonian for the thermofield double state, explained its properties, and showed that given the above form of a Hamiltonian we can derive such a thermofield double state by practical variational means. 
The AdS/CFT duality, being a holographic duality, is particularly interesting in quantum gravitational discussions. However, the duality in its practical realisations offers insight for the case of conformal field theories, and therefore it is interesting to describe quantum states in which the two connected regions in the above description of thermofield double states are indeed presented as conformal (field) theories. 
We therefore re-write the TFD-Hamiltonian as
\begin{equation}
\begin{array}{c}
H_{TFD}=\sum_{\alpha}\lambda_{\alpha}d_{\alpha}^{\dagger}d_{\alpha}+\sum_{i}c_{i}d_{i}^{\dagger}d_{i}\\
\\
d_{\alpha}=e^{-\beta (H_{L}^{0}+H_{R}^{0})/4}(J_{\alpha}^{L}-\Theta J_{\alpha}^{R\dagger}\Theta^{-1})e^{\beta(H_{L}^{0}+H_{R}^{0})/4}\\
\\
\end{array}
\end{equation}
where $J_{\alpha}^{L,R}$ are the generators of the conformal algebra in the left and right theories and $d_{i}$ is the operator defined above but this time for a primary operator $\mathcal{O}_{i}$. It has been shown in [26] that it is not necessary to also include the $d$ operators with conformal descendants in the TFD hamiltonian. It is interesting to note that terms proportional to $\lambda_{\alpha}$ make the Hamiltonian non-local at all scales. 
With this construction, and with Susskind's original observation that there is a bi-directional relation between entanglement as expressed by the EPR gedankenexperiment and ER-wormholes, it has been noticed by various authors that entanglement by itself, as understood up to now, is not sufficient to properly account for the type of correlation one requires in the black hole context [27], [28]. The thermofield double state expressed above can be understood as an Euclidean path integral over an interval of length of $\beta/2$ in Euclidean time. If we look at the holographic dual, an uncharged black hole represents a bulk saddle point for the Euclidean boundary conditions. If this saddle point is dominant, when we continue to the Lorentzian picture the dual of such a state would be a maximally extended black hole. With this observation in mind, it was Susskind to notice that entanglement can be considered as a connection through the bulk geometry. If we use the operators in the two copies of the thermofield double state, namely in the two copies of the system, to form a two-point function of the form 
\begin{equation}
\Bracket{\mathcal{O}_{L}\mathcal{O}_{R}}
\end{equation}
we can find such functions that have non-zero values in this state. Therefore the bulk connection between the two sides can allow for interactions to propagate from one boundary to the other, while in the boundary quantum field theory, the two point functions are non-vanishing due to entanglement, although no interaction can propagate in the boundary region between them. In ref. [24] it has been asked what states in the combined Hilbert space of the left and right states can have such a wormhole description. Given that no entanglement would not be possible as that would cancel out the two point correlation function, what alternatives could there be?
The ER=EPR conjectures states that entanglement should be identified with the existence of a wormhole. However, by means of the Eigenvalue thermalisation hypothesis, it has been argued that the local correlations in a typically entangled state are weak and therefore are not easily associated to a semiclassical wormhole in the bulk [33]. There have been found examples of special states corresponding to long semiclassical wormholes with weak local correlations that allowed smooth wormholes. It appears that the difficulty emerges from understanding when the semiclassical nature of wormholes becomes manifest and what other type of (quantum) wormhole could exist? Whether deformations of the usual thermofield double states allow for semiclassical wormhole interpretations has been analysed in various articles in the literature [34], [35]. 
Indeed, it has been shown that the usual entanglement would not always be sufficient, and therefore, new alternatives are particularly important. Such alternatives are being discussed in what follows. 
 
\section{speed of light and speed of information retrieval}
As stated in the introduction, the conventional way in which information is obtained in a remote region of spacetime, remote in the sense that it is separated (presumably in a timelike manner) from the region where the phenomenon producing that information occurred, is by light signals, or by massless gauge fields, otherwise known as gauge connections, on fibre bundles of various types. In general we consider that in order to have access to the information occurring in some region, we have to follow the gauge connection (say a light beam) and extract the information in the points where that light beam is received. This is the basis of special relativity and stays at the foundations of concepts like causality, the causal cone, or locality. We underline the distinction between a local phenomenon and a causal phenomenon, hence the two different names used. This is not the only, nor even the usual way in which information is extracted in similar problems by mathematicians. If we want to know the global result of the transport across fibres of a fibre bundle with a connection, and infer properties of the connections from it, following the fibre connection is probably the most mundane, unimaginative way to follow. We could infer the same information by instead following the possible ways in which our initial bundle structure can be mapped into some benchmark bundle that has well known characteristics. Using that knowledge and the knowledge of the maps involved in the transformation, we may be able to predict the results of propagating information across fibres in a bundle, without actually having to propagate them practically. These types of "tricks" are common in mathematics. One has to be a poor mathematician indeed to perform only the brute force approaches to obtain relevant information, and following a connection on a bundle is probably a very brute force approach, which Nature doesn't seem to use in some very interesting cases (like, for example, in black hole physics). At the same time, we have a limit on information propagation, which is the well known speed of light in a vacuum, a relativistic invariant that is preserved across transformations of reference choices. The question of whether there is a limit on the speed of information retrieval and what that speed might be is an open question, and the understanding of the scrambling time of a black hole is one important aspect of this question [14], [15], [16]. Another important aspect is understanding possible distinct mechanisms for information retrieval. One such mechanism is presented in this article. 
We want to describe information "retrieval" (I tried to use another word, but it must be something that is not related to "transmission") from the global, non-separable structure of the maps between states. This will be some form of entanglement entropy, but not the usual one. I discussed the concept of entanglement entropy for uncertainty in topology in my article [11]. There, an uncertainty in the topology of gauge field spaces was considered, due to the inability of some arbitrary topological invariant to perfectly clarify all possible topological structures at once. I gave several examples of how this can happen in that paper of mine and the subsequent ones [12], [13]. However, I soon realised that this type of added complexity does not solve the black hole paradox by itself, because, gauge space or not, I was still analysing the situation from a state-oriented point of view, without regard for the structure of the maps between states. A similar entropy can be defined globally for the maps between states, and that entanglement entropy massively enhances the information complexity, and ultimately gives a better definition of the notion of complexity. While enhancing it in a quantitative sense, it also simplifies the analysis in various ways. We have several algebraic results that can offer us ways to explore those topological and geometrical features constructed by maps.

\par
It is important to underline that we will always consider the restriction that the transformations induced are unitary, in order to preserve the defining characteristic of quantum mechanics. This is not affecting the general discussion, remaining for this restriction to be re-introduced at the end of the construction. Unitary maps clearly have all the structures required here, including non-trivial topology, geometry, curvature, etc. or else we wouldn't have a unitary description of gauge fields to begin with. 
As noted by Grothendieck, many properties of a topological space X can be formulated in terms of invariant properties of sheaf categories. 
In particular, as an intuition for what is to come, we can define a generalised tensor product which will give us a new notion of non-cartesianity which expands what we mean by "entanglement". But more reading must come before that. 
One construction we will need will be that of a topos. Grothendieck introduced it in order to make sense of concepts related to geometry and topology in areas that are not immediately amenable to such an approach. Toposes are generalisations of categories of sheaves of sets where the topological space $X$ is replaced by a so called "site" which is basically a pairing $(\mathcal{C}, J)$ made out of a "small" category $\mathcal{C}$ and some notion of covering on that site called $J$. Taking the sheaves over those objects leads to the commuting square 
\begin{equation*}
  \xymatrix@R+2em@C+2em{
  X \ar[r]^-f \ar[d]_-g & Sh(X) \ar[d]^-h \\
  (\mathcal{C}, J) \ar[r]_-k & Sh(\mathcal{C},J)
  }
 \end{equation*}
The Grothendieck topology on a category produces a categorification of the covering of an open set of a topological space with a given family of open subsets. To define it we should understand the concept of a sieve. Basically, it is simply a way in which an object in a category is the result of the mapping of all other objects of the category. The sieve is basically all the arrows that link the other objects of the category with our final object, let's call him $c$. For the construction to actually be a sieve we need some rule that makes it possible to compose such arrows, basically to have a well defined structure such that, given a sieve $S$
\begin{equation}
f\in S \Rightarrow f\circ g \in S
\end{equation}
With this, we start having a concept of factorisation which already looks slightly more general than what we were used to in the basic quantum mechanics usually employed in TFD descriptions. We may say that a sieve $S$ is generated by a pre-sieve $P$ on an object $c$ if it will be the smallest sieve containing it, i.e. if it is the set of all maps towards $c$ which factor through an arrow in $P$. 
The Grothendieck topology is defined for a small category $\mathcal{C}$ is a function $J$ linking each object of our small category to a collection of sieves on $c$ with the properties that (1) (maximality axiom) the maximal sieve $\mathcal{M}_{c}=\{f|cod(f)=c\}$ is in $J(c)$, (2) (stability axiom) if $S\in J(c)$, then also $f^{*}(S)\in J(d)$ given any arrow $f:d\rightarrow c$, and (3) (transitivity axiom) if $S\in J(c)$ and $R$ is any sieve on $c$ such that $f^{*}(R)\in J(d)$ for all $f: d \rightarrow c \in S$ then $R\in J(c)$. 
We call site a pair $(\mathcal{C}, J)$ where $\mathcal{C}$ is a category and $J$ is a Grothendieck topology on $\mathcal{C}$. 
The sieves $S$ belonging to $J(c)$ are said to be $J$-covering. A site $(\mathcal{C},J)$ is said to be small generated if $\mathcal{C}$ is locally small and has a small $J$-dense subcategory i.e. $\mathcal{C}$ admits a $J$-covering sieve such that the arrows from objects in a small $J$-dense subcategory compose with each other leading to compositions in the same small $J$-dense subcategory. 
Now we define the pre-sheaves on a small category $\mathcal{C}$ as the functor from the opposite of the category (the category with all the map arrows reversed) to the Set category. We take a sieve on the object $c$ of $\mathcal{C}$, calling it $S$ and define a matching family for $S$ of elements in $P$ as a function assigning to each arrow $f:d\rightarrow c$ in $S$ an element of $P(d)$ such that 
\begin{equation}
P(g)(x_{f})=x_{f\circ g}, \; \forall g: e\rightarrow d
\end{equation}
and we define an amalgamation for such a family as a single element $x\in P(c)$ such that $P(f)(x)=x_{f}$, $\forall f\in S$. 
A sheaf is a pre-sheaf for which for any matching family and any J-covering on any object of $\mathcal{C}$ has a unique amalgamation. With these notions, a Grothendieck category is any category equivalent to the category of sheaves on a small site. 
Grothendieck toposes satisfy all categorical properties. The notion of morphism for such a category is the geometric morphism. 
Let us follow again the arrows and connect them to the problem of complexity and information. The main advantage of Grothendieck's construction including his topology is that basically we can construct a topology and a geometry on almost anything. This is of course an exaggeration, but not by much. As said before, we try to map quantum states into quantum states, and we find out that the maps have a "life" of their own, with topology, geometry, and even dynamics (as a generalisation of the Gauss-Mannin differential equations, also due to Grothendieck). Such structures will also have a notion of cartesianity and of obstruction to cartesianity which will lead to a new notion of entanglement and entanglement entropy. About that shortly. 
\par First let us follow the maps: 
the Grothendieck topology is a function that assigns to each object of our category a collection of sieves from the collection of maps in our category. Loosely speaking we will obtain a sieve of maps leading from all other states around to one particular state we have chosen. This is (at least locally) our "terminal" state, as far as we are concerned at one specific step in the procedure. These sieves have various properties, the most important one being that composing two maps of the same sieve will bring us to the same sieve, hence we have a well defined composition rule on this structure. If we reverse all the maps of our category, and we define a functor that brings us to a set category we can obtain unique results as elements of the set through this functor relating arrows from our sieve to elements of our sets. 
Therefore we obtained a consistent way in which we can discuss both the states and the maps, with enough structure to define geometric and topological information on the maps. 
\par Now comes the interesting part: we want a generalisation of the tensor product. This is very important, because, we know that the main aspect of quantum mechanics we all like, entanglement, is the result of going from a cartesian product to a tensor product defined by means of linear transformations, such that pairing two object via such a product sometimes creates non-separable structures, hence entanglement. We also have such a concept on the maps. 
In general let us try to define this generalisation for the categories introduced by Grothendieck and applied here to our maps between states. 
Let us take a small category $\mathcal{C}$ and another locally small category $\Omega$. The functor connecting them will be $A:\mathcal{C}\rightarrow \Omega$. We can define the adjunction 
\begin{equation}
L_{A}:[\mathcal{C}^{op},Set]\leftrightarrows \Omega : R_{A}
\end{equation}
Given two objects $o\in Ob(\Omega)$ and $c\in Ob(\mathcal{C})$ we can define the right adjoint as the following Homomorphism 
\begin{equation}
R_{A}(o)(c)=Hom_{\Omega}(A(c), o)
\end{equation}
and the left adjoint as the colimit 
\begin{equation}
L_{A}(P)=colim(A\circ \pi_{P})
\end{equation}
where $L_{A}:[\mathcal{C}^{op}, Set]\rightarrow \Omega$ and 
$\pi_{P}$ is the canonical projection functor $\int P\rightarrow \mathcal{C}$. 
The colimit is defined in a category, such that it leads us to the co-classifying space for morphisms in that category. That means that in a diagram of that category we obtain as a co-limit the object generated by linking together the objects of the diagram by means of the rules defined by the morphisms of the diagram. The diagram in this case will be a co-equaliser 
\begin{equation}
\coprod_{c\in\mathcal{C}, p\in P(c), u:c'\rightarrow c}A(c')\rightrightarrows\coprod_{c\in\mathcal{C}, p\in P(c)}A(c)\xrightarrow{\phi}L_{A}(P)
\end{equation}
which can be regarded as an equivalence relation in a quotient prescription for objects in our category. 
The first upper arrow is $\theta(c,p,u,x)=(c', P(u)(p),x)$ and the lower arrow is $\tau(c,p,u,x)=(c,p,A(u)(x))$. 
Now we may write 
\begin{equation}
L_{A} \sim *\otimes_{\mathcal{C}}A:[\mathcal{C}^{op},Set]\rightarrow \Omega
\end{equation}
which brings us to the equivalent form of the co-equalizer 
\begin{equation}
\coprod_{c, c' \in\mathcal{C}}P(c)\times Hom_{\mathcal{C}}(c',c)\times A(c')\rightrightarrows\coprod_{c,\in\mathcal{C}}P(c)\times A(c)\xrightarrow{\phi}P\otimes_{\mathcal{C}}A
\end{equation}
and here we are, obtaining our generalised tensor product
\begin{equation}
P\otimes_{\mathcal{C}}A \cong A\otimes_{\mathcal{C}^{op}}P
\end{equation}
This is a general result by Grothendieck, repeated in various aspects across category theory. Let us see what exactly does it provide us with in our problem and how it brings us to a more general concept of entanglement entropy (aka non-cartesianity of pairing of spaces of maps inside categories).
We get a pairing between the results of the functor which encodes a projection from the category of elements of the Set to which we mapped our initial category and the locally small co-complete category $\Omega$. While defining this we used an equivalence type relation in the quotient, generating basically equivalence classes linked by the very maps we defined. This makes the tensor product we defined here a result of a fibre bundle over the maps and hence a clear global obstruction against cartesianity. Therefore there are several ways in which we can define this as some form of entanglement entropy. The entanglement entropy will count the cardinality of the classes of equivalence of the co-equalizer for the maps we defined, and will inherit its structure from the original category $\mathcal{C}$. At the same time, it results from the new, generalised tensor product defined for the maps we are considering. 
\par In general in quantum mechanics we define the tensor product as a result of the linear transformations of the states we consider, but we never consider any additional structure associated to them. I think this is a deficiency for various reasons. In a space with curvature, those maps inherit the geometrical properties of the underlying space, and produce the relativisation of the vacuum, as we may have learned from both Unruh and Hawking. But this is by far not the only aspect that we encounter. In situations in which the underlying spacetime has a horizon (as is the case of black holes) we encounter a change in the open sets defining the topology, at least in one direction (out of the black hole). The maps must adapt to that as well, and the generalised Grothendieck construction will then allow us to map the resulting bundles in a consistent way, bundles describing topological properties of maps instead of simple flat space transformations. So, here we did generalise the tensor product to a construction that takes into account the categorical nature of the quantum information, namely combining the quantum states with the potential mappings in a more general way, including various types of possible patches that must occur near or around the black hole horizon. We discovered a new, more intricate form of entanglement entropy and even of entanglement, when we analyse the category formed out of objects and their mappings, and finally, we found a new method of extracting information from the black hole surroundings. The notion of complexity I am proposing here is one that takes into account all the possible global correlations of the maps in the category, apart of the simple transformations considered up to now between the objects. 
\par To make sure this is clear, and to underline that a similar phenomenon is well known for the case of flat connections in usual gauge theories (say, the Chern Simons construction), I will construct geometric morphisms as flat functors, in a categorical sense, again, of course, following closely the constructions by Grothendieck (I hope it is clear I didn't invent any of the mathematics here, and I cannot claim I understand it in all details, all I say is that what I understand seems to have highly non-trivial implications for the problem of the black hole paradox). 
We call a functor $A:\mathcal{C}\rightarrow \Omega$ that connects a small category $\mathcal{C}$ to a locally small topos $\Omega$ flat if the limits induced by 
\begin{equation}
*\otimes_{\mathcal{C}}A:[\mathcal{C}^{op},Set]\rightarrow \Omega
\end{equation}
will be finite. The subcategory of flat functors of $[\mathcal{C},\Omega]$ will be denoted $Flat(\mathcal{C},\Omega)$. The evolution of connections on fibre bundles defined on the fibres above such categories will induce the equivalent of gauge "fields" with "strengths" that are equivalent to non-flat contributions, as well as topological terms. A very interesting generalisation indeed, in which the states and the maps cannot be completely separated of each other and in which maps can have various structures adding to the complexity. The complexity is usually defined with respect to a given reference state, and the construction of a set of transformations which will strongly depend on the underlying geometry, topology, etc. but will develop also their own categorical structures which their own topology, geometry, etc. and the connections associated to them. Fortunately this does not increase the difficulty of the problem, instead, it allows us to infer new results regarding the information we desire, from all the categorical structures emerging on both objects and maps. 
\section{statements}
On behalf of all authors, the corresponding author states that there is no conflict of interest.

\end{document}